\documentclass[twocolumn,superscriptaddress,aps,prl,floatfix]{revtex4-1}
    \usepackage[version=3]{mhchem}
    \usepackage[utf8]{inputenc}
    \usepackage[english]{babel}
    \usepackage{setspace}
    \usepackage{graphicx}
    \usepackage{color}
    \usepackage{natbib}
    \usepackage{colortbl}
    \usepackage{enumitem}
\begin{document}
%\doublespacing
%%%%
%%%%---------------------------------------------------------------------------------------------------------
\title{Static disorder and structural correlations in the low temperature phase of lithium imide}
\author{Giacomo Miceli}
\affiliation{Dept. of Materials Science, Universit\`{a} di Milano-Bicocca, via R. Cozzi 53, I-20125 Milano, Italy}
\affiliation{Computational Science, DCHAB, ETH Zurich, USI Campus, via G. Buffi 13, CH-6900 Lugano, Switzerland}
\author{Michele Ceriotti}
\affiliation{Computational Science, DCHAB, ETH Zurich, USI Campus, via G. Buffi 13, CH-6900 Lugano, Switzerland}
\author{Marco Bernasconi}
\affiliation{Dept. of Materials Science, Universit\`{a} di Milano-Bicocca, via R. Cozzi 53, I-20125 Milano, Italy}
\author{Michele Parrinello}
\affiliation{Computational Science, DCHAB, ETH Zurich, USI Campus, via G. Buffi 13, CH-6900 Lugano, Switzerland}

\begin{abstract}
Based on \textit{ab-initio} molecular dynamics simulations, we investigate the low temperature
crystal structure of Li$_2$NH which in spite of its great interest as H-storage material is still
matter of debate. The dynamical simulations reveal a precise correlation in the
fractional occupation of Li sites  which leads average atomic positions in excellent
agreement with diffraction data and solves inconsistencies of previous proposals.
\end{abstract}

\maketitle
%---------------------------------------------------TEXT---------------------------------------------------

Lithium amide (LiNH$_2$) and imide (Li$_2$NH) have been
extensively studied in recent years as promising materials
for hydrogen storage\cite{ping1,ping2,gregory,mtoday,orimo}. 
Hydrogen release occurs in the
mixture LiNH$_2$/LiH via a reversible solid state decomposition
reaction into lithium imide and molecular hydrogen,
LiNH$_2$ + LiH $\rightarrow$ Li$_2$NH + H$_2$. 
The typical operating
temperature for this system is around 280 $^o$C, which
is probably too high for on-board applications. Nevertheless,
the amide/imide system is under deep scrutiny
since it represents a prototypical, relatively simple system,
which could shed light on the mechanisms of reversible
H-release in the more complex, and technologically more 
promising, reactive hydrides made of mixtures of
amide, borohidrides and/or alanates (e.g. LiNH$_2$/LiBH$_4$
or LiNH$_2$/NaAlH$_4$)\cite{orimo,mtoday}.  
The search for better performing materials 
in this class would greatly benefit from a microscopic knowledge of the decomposition
mechanism which  requires in turn a detailed description of the 
crystalline phases involved. For several 
complex hydrides the structure of the  phases undergoing the 
dehydrogenation/rehydrogenation process
 is still not fully resolved. This is the case for
instance of the most studied 
sodium alanate  (NaAlH$_4$) for which Raman spectroscopic data\cite{yukawa} and 
ab-initio simulations\cite{wood-marz09prl} very recently suggested the existence of 
a new high temperature phase  which is expected to  mediate  the decomposition reaction  
in place of the low temperature $\alpha$-phase considered so far. 
The structure of Li$_2$NH as well is still a matter of debate.  Structural refinement
from neutron and x-rays diffraction data reveals a structure of Li$_2$NH with fractional
occupation. In spite of a substantial amount of experimental and theoretical
investigation, the problem of the actual local structure which yields this long-range
disorder is still unsettled.
Based on ab-initio simulations, we have identified  a model for the  local 
structure of the low-temperature phase of lithium imide (Li$_2$NH) which 
solves inconsistencies of previous proposals and fully agree with experimental data available.

\begin{figure}[!ht]
\centerline{\includegraphics[width=0.8\columnwidth]{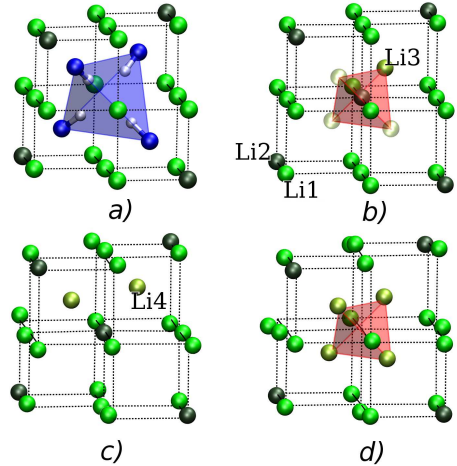}}\vspace{-2ex}
\caption{ \textit{(a)} Tetrahedral arrangement of \ce{NH} groups around a Li$^+$ vacancy in the LT phase of
 Li$_2$NH. \textit{(b)} Li ions are distributed on the three sites independent by symmetry labeled in panel.
The interstitial Li3 sites have fractional  occupation. Sites which are left
empty in the $Ima2$ model of Ref.~\cite{herbst} are shown as transparent spheres.
 \textit{(c)}  Occupied Li4 octahedral sites obtained upon relaxation
of the $Ima2$ model.
\textit{(d)} Our proposed correlation in the occupation of Li3 and Li2 sites leads to a
tetrahedral cluster of four interstitials ions at Li3 sites around a vacancy at Li2 site.
}
\label{fig:tetra}
\end{figure}

Differential thermal analysis and NMR measurements\cite{imideOD1,imideOD2}
in the late 60's revealed a reversible phase transition
at 356 K between an unknown low temperature (LT) structure and a high temperature antifluorite phase of Li$_2$NH.
Structure refinement from x-ray and neutron diffraction measurements on
deuterated imide (Li$_2$ND) have been published only very recently\cite{balogh}.
The high temperature phase yields a diffraction pattern consistent with an anti-fluorite structure,
in which hydrogen atoms occupy the $192l$ positions of the $Fm\bar{3}m$ space group.
At low temperature (100-300 K) the diffraction data were best fitted by a cubic crystal with
$Fd\bar{3}m$  space group. The LT crystal was described as a superstructure of the antifluorite phase
in which one out of 8 Li atoms is displaced to an interstitial site giving rise to Li vacancies arranged in
an ordered manner and tetrahedrally coordinated to four NH groups (Fig.~\ref{fig:tetra}(a)).
The structure is stabilized by electrostatic
interaction of H$^{\delta+}$  pointing towards the Li$^+$ vacancy which is formally a negatively charged site.
Similar tetrahedral arrangements  of four NH groups are present also in other imides, such as \ce{Li2Mg(NH)2} \cite{rijss}.
In the LT phase, H and N atoms occupy $32e$ sites.
Li atoms are distributed over three different sites (Fig.~\ref{fig:tetra}(c)):
$48f$ (Li1), $8a$ (Li2) and $32e$ (Li3). The Li3 site corresponds to a displacement along 
$\left<111\right>$ directions from  the octahedral position 
Li4 (sites 16d of the $Fd\bar{3}m$ space group, Fig. 1(c)).
According to the Rietveld refinement,
the latter has a fractional occupation of about $1/3$ at low temperatures,
which only at room temperature gets closer to  the value required by stoichiometry ($1/4$).

Diffraction data, however, do not provide information on correlations among the occupation of sites 32e which
are mandatory to get full insight on the structural and dynamical properties of the system.
To this aim, an attempt was made by Herbst and Hector to model  the  $Fd\bar{3}m$ crystal
by fully occupying selected 32e sites \cite{balogh,herbst,hector}.
In the structure proposed, a single Li$^+$ ion is present  around each Li2 site, resulting in  a $Ima2$ symmetry (Fig.~\ref{fig:tetra}(b)).
The distance between Li2 and Li3 sites is, however, very small (1.53 $\rm\AA$) and
upon geometry optimization performed by \textit{ab-initio} calculations\cite{herbst},  Li interstitials
are displaced from Li3 sites into octahedral Li4 positions (sites $16d$ of  the  $Fd\bar{3}m$
space group) with a longer Li4-Li2 distance of 2.3 $\rm\AA$ (Fig.~\ref{fig:tetra}(c)).
The locally relaxed structure obtained by Hector and Herbst~\cite{herbst} has $Imma$ symmetry and will
be referred to hereafter as the HH phase. The positions of Li interstitials at Li4 sites are, however, incompatible with
Rietveld refinement. Moreover, the theoretical formation energy of this phase is too high when compared to experimental data.
Thus, other structures have been proposed on the basis of theoretical calculations, with Li atoms arranged in
an ordered manner and a slightly lower formation energy\cite{wolverton,ceder}. 
However, their equilibrium lattice parameters are inconsistent with the space group symmetry
inferred experimentally.
On the other hand, the local instability of the ideal
antifluorite structure of Li ions
at low temperature was confirmed by
\textit{ab-initio} molecular dynamics simulations at 300 K which
showed the spontaneous formation of Li Frenkel pairs\cite{mice+2010-jpcc}
and strong distorsions of the Li sublattice\cite{sebastiani}.

Based on \textit{ab-initio} simulations, we propose a structure for the LT phase which solves the
problems mentioned above by introducing vacancies on partially occupied Li2 sites with a precise correlation with the
occupation of Li3 interstitial sites. The solution of the puzzle came from the analysis of 
{\em ab-initio} molecular dynamics trajectories.

We started our analysis by performing ab-initio molecular dynamics simulations on the HH structure.
A $\sqrt{2}\times 1\times\sqrt{2}$, 128-atoms supercell corresponding to two $Imma$ unit cells was used.
We performed Born-Oppenheimer molecular dynamics simulations within Density Functional Theory (DFT)
with gradient corrected exchange and correlation functional\cite{PBE} as implemented in the CPMD~\cite{CPMD} package.
Ultrasoft\cite{vanderbilt} and Goedecker-type\cite{GTH1} pseudopotentials were used,respectively,
for N and H atoms and for Li with three valence electrons. Kohn-Sham orbitals were expanded
in plane waves up to a kinetic-energy cutoff of 50 Ry. Brillouin Zone (BZ) integration was restricted to the
 $\Gamma$ point only. A time step of 0.6 fs was used and a constant temperature of
300K was enforced by an optimal-sampling generalized Langevin equation thermostat\cite{colored-jctc}.
Equilibrium geometries of relevant structures emerged from the dynamical simulations were optimized
with special k-points meshes and the Quantum-Espresso suite of programs \cite{Quantum-espresso}.
Activation energies for diffusion processes discussed below were obtained by
Nudged Elastic Band optimizations\cite{NEB2}.

In the dynamical simulations, 15 ps long, we observed that constitutional vacancies coordinated by NH groups are very stable,
the latter  performing only small librations around their equilibrium position. Instead,
Li interstitials occupying Li4 positions are very mobile and we observed several jumps between adjacent sites,
taking place via an exchange mechanism with one of the Li1 atoms (cf. Fig. S1 in supplementary materials below).
The activation energy for diffusion of interstitials turns out to be as low as 0.13 eV.
Moreover, we verified that many different
arrangements of interstitial atoms in Li4 sites have total energies within few meV.
At finite temperature,  we would thus expect a disordered arrangement of interstitial atoms in the Li4 sites,
leading to a fractional occupation of 1/2 and an overall  $Fd\bar{3}m$ space group.

%------------------------figure2
\begin{figure}[!ht]
\includegraphics[width=1.0\columnwidth]{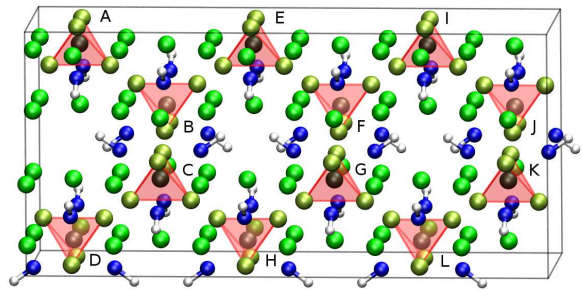}
\caption{ \textit{a)} The  192 atoms supercell is shown ($3\times 1\times 1$ HH unit cells). The labeling
indicates all possible sites where the four  Li2 vacancy -- Li3 interstitials
clusters can be distributed inside the supercell.
We found that the lowest-energy arrangements are $ACFL$ (-30  meV/f.u.,
see also Fig.~S2 below) and  $AEGK$ (-29).
Energies are given relative to the relaxed HH structure.
Configurations where the tetrahedra are unevenly distributed such as
$AEIC$ (-17) or $ACEG$ (4) are higher in energy, and those where
nearest-neighbor sites are occupied simultaneously are even less stable.
}
\label{fig:311}
\vspace{-2ex}
\end{figure}

However, the positions of Li4 still differ from those of Li3 obtained from Rietveld refinement.
Moreover, the low diffusion barriers found would assign a superionic
character to Li$_2$NH even at room temperature for which no experimental evidence has been provided so far.

The solution to these discrepancies came from a
closer inspection of the molecular dynamics trajectory, which revealed that
as soon as three interstitial atoms happen to be all nearest neighbors to an occupied Li2 site,
the latter atom is displaced from its position.
A tetrahedral cluster is formed with a vacant Li2 site at its center (Fig.~\ref{fig:tetra}(d)).
These clusters remain stable thereafter and prevent diffusion of Li interstitials.
The interstitial Li now occupies the Li3 $32e$ sites and a vacancy is introduced at site Li2 in such a way that
the very short Li3-Li2 distance of the unrelaxed HH model is removed.
%%%%%%%

Having observed the propensity of Li atoms to form these rather stable tetrahedral structure,
we propose a new stable structure in excellent agreement with diffraction data, as follows.
At T=0 we expect that all the Li interstitials are involved in the
formation of tetrahedral clusters.
The smallest supercell compatible with this requirement and consistent
with the stoichiometry is formed by 3$\times$1$\times$1 HH unit cells (192 atoms).
In such supercell there will be four vacancy-interstitials clusters, which can be distributed over twelve Li2 sites
(see Figure~\ref{fig:311}).
We considered all the possible arrangements of the vacancy-interstitials tetrahedra independent by symmetry
excluding those where two first-neighbor Li3 sites are occupied.

Provided that
the clusters are evenly distributed in the cell to balance the negatively-charged constitutional
vacancies, the spread in the total energy
of the different configurations is well within thermal  energy at room temperature (see Fig.~\ref{fig:311}).
We repeated our analysis on a larger $3\times 1\times 3$ supercell finding similar results.
Such degeneracy allows for long-range disorder in the arrangement of tetrahedra, so that
a diffraction pattern consistent with a higher-symmetry $Fd\bar{3}m$,
with a fractional occupation of sites Li3 (1/3) and Li2 (2/3) is to be expected.
These fractional occupations are compatible with the stoichiometry and solve the
inconsistency in the model of Ref.~\cite{balogh}.

By averaging over possible configurations of our 192-atoms supercell we obtained
the symmetry-adapted average positions reported in Table~\ref{tab:positions}
and the relative mean square displacements which would correspond to a static contribution
to the Debye-Waller factor. The presence of clusters generates long-range distortions
in the Li1 lattice and in the orientation of H atoms around a constitutional vacancy,
resulting in a large mean square displacement from the average position for these species.
The thermal Debye-Waller factors as a function of temperature have been computed from harmonic
phonons (at the supercell $\Gamma$-point) and are given in Fig. S3 in supplementary materials below and at $T=100$~K in Table~\ref{tab:positions}.

The agreement with the experimental positions from Rietveld refinement is excellent.
The diffraction pattern of our proposed structure compares well with that deduced from the experimental positions and
fractional occupation (Fig. S4 in supplementary materials below).
\begin{table}[!h]
\centering
\begin{tabular}{lcc}
\hline \hline
Site  & $x$ (exp. Ref~\cite{balogh}) & $U_{iso}$ static/thermal  \\
\hline
N   $32e$ f=1    &  0.2408 (0.2418) & 0.34/0.50  \\
D   $32e$ f=1    &  0.2982 (0.2976) & 1.58/2.00  \\
Li1 $48f$ f=1    &  0.3755 (0.3734) & 2.18/0.93  \\
Li2 $8a$  f=2/3  &  --              & 0.25/0.87  \\
Li3 $32e$ f=1/3  &  0.0367 (0.0376) & 0.22/0.91  \\
\hline \hline
\end{tabular}
\caption{Structural parameters of  the best-fit $Fd\bar{3}m$ structure
corresponding to the $ACFL$ arrangement of tetrahedral clusters in a $3\times 1\times 1$
supercell (see Fig.~\ref{fig:311}).
The Debye-Waller factor ($10^{-2}$\AA$^2$) is the sum of a contribution due to
 static disorder, and a thermal contribution calculated at $T=100$~K for Li$_2$ND.
The experimental data from Rietveld refinement of neutron diffraction
patterns at $100$~K from Ref.~\cite{balogh} are given in parentheses.
}
\label{tab:positions}
\end{table}

The formation enthalpy of the $ACFL$ configuration (see Fig.~\ref{fig:311}) is lower 
(-30 meV f.u.$^{-1}$) than that of the HH structure.   
Considering also zero-point energy and the harmonic contribution to the total free energy,
our structure is still more stable than HH's (-18 meV f.u.$^{-1}$ at 298 K) and 
the structure proposed in Ref. \cite{wolverton} (-3 meV f.u.$^{-1}$).
Our structure is  slightly less stable (by 22 meV f.u.$^{-1}$) than that suggested in ref.\cite{ceder} 
which however does not correspond to the experimental space group. 
However,  uncertainties in the relative energies of the different phases 
are expected due to current approximations to
the exchange and correlation functional as occurs for instance in other materials we have recently studied\cite{sosso}.
Moreover, entropic contributions due to disorder
might compete with small enthalpy differences of ordered models and need to be
taken into account before assigning the most stable phase at the DFT level.
Because of all the above limitations, our aim was not to find the lowest 
possible ordered structure on the potential energy
surface as pursued in Refs. \cite{wolverton,ceder} but to refine a 
disordered structure which was a local minimum in DFT in such a way
that is compatible with the experiments.

We verified the stability of the tetrahedral clusters by performing $20$~ps of molecular dynamics at $300$~K
using the most stable arrangement $ACFL$. The vacancy-interstitials clusters
do not break nor constitutional vacancies diffuse on the simulation time scale.
The computed energy barrier to break a tetrahedral cluster by diffusion of a Li3 ion is indeed 0.5~eV,
as obtained by NEB optimization, resulting in a low  mobility of Li atoms in the LT phase
(Li3 diffuses by an exchange mechanism similar to that of Li4, see Fig. S1 in supplementary materials below).
The energy cost to break a cluster in the $ACFL$ structure,
via  the reaction $4\,\text{Li3}\rightarrow \text{Li2}+3\,\text{Li4}$ is $\Delta$E=0.20~eV.
This energy depends on the final configuration of the interstitials, and on the number of
clusters which are simultaneously broken. For instance, the difference in energy with the HH structure - with all
the clusters broken - amounts to $\Delta$E=0.36~eV per cluster.

We can compute the change in configurational entropy due to breaking of tetrahedra
and formation of these interstitial Li4 atoms, by assuming a perfectly random
distribution of tetrahedra and occupied Li4 interstitial sites (see supplementary materials below) which
finally yields a concentration of broken tetrahedra of $x \sim \sqrt[3]{4}/3\,exp[-\Delta \textrm{E}_0 /3 k_B T]$.
This corresponds to a concentration at 300~K between $\sim$6\% and $\sim$0.5\%, depending
on the estimate taken for $\Delta$E$_0$.
In spite of this large concentration of Li4 the diffusivity is still low
because most of the percolating paths for Li4 interstitials are suppressed by the presence of
Li3 atoms fixed in tetrahedra.

We speculate that the possible coexistence of partially occupied
Li3 and Li4 sites might contribute to the change with temperature of the
occupation of Li3 sites emerged from the Rietveld analysis \cite{balogh}.
We encourage new refinement of the diffraction data starting from our structural parameters in
Table~\ref{tab:positions}, eventually allowing for a (small) partial occupation of sites Li4.

In summary, by combining \textit{ab-initio} molecular dynamics and structural optimization, we
have provided a full description of the structure of the LT phase of Li$_2$NH.
Atomic positions coincide with those inferred from diffraction data but
partial occupations of Li sites have a strong nearest neighbor
correlation which solve the inconsistencies raised by previous proposals.
This work represents an exemplary demonstration of how dynamical simulations
can provide crucial insights to fully resolve the structure of systems with partial
disorder, complementing experimental diffraction data.

We gratefully thank W. I. F. David for discussion and information.

%Merlin.mbs v4.21 2009-07-09.
%
%----------------------------------------------------------------------------------------------------------
\clearpage
\onecolumngrid
{\begin{center} \large \bfseries  Supplementary materials  \end{center}}
\renewcommand{\thepage}{S\arabic{page}}
\renewcommand{\thefigure}{S\arabic{figure}}
\setcounter{figure}{0}
\setcounter{page}{1}
\section*{Computational Details}
We started our analysis by performing ab-initio molecular dynamics simulations on the HH structure.
A $\sqrt{2}\times 1\times\sqrt{2}$, 128-atoms supercell corresponding to two $Imma$ unit cells was used.
We performed Born-Oppenheimer molecular dynamics simulations within Density Functional Theory
with gradient corrected exchange and correlation functional\cite{PBE} as implemented in the CPMD~\cite{CPMD} package.
Ultrasoft\cite{vanderbilt} and Goedecker-type\cite{GTH1} pseudopotentials were used,respectively, 
for N and H atoms and for Li with three valence electrons. Kohn-Sham orbitals were expanded 
in plane waves up to a kinetic-energy cutoff of 50 Ry. Brillouin Zone (BZ) integration was restricted to the
 $\Gamma$ point only. A time step of 0.6 fs was used and a constant temperature of
300K was enforced by an optimal-sampling generalized Langevin equation thermostat\cite{colored-jctc}.
Equilibrium geometries of relevant structures emerged from the dynamical simulations were optimized
with special k-points meshes and the Quantum-Espresso suite of programs \cite{Quantum-espresso}.
Activation energies for diffusion processes discussed below were obtained by 
Nudged Elastic Band optimizations \cite{NEB2}.

\section*{Mechanism of diffusion of Li interstitials}
\begin{figure*}[!h]
        \includegraphics[width=0.35\textwidth]{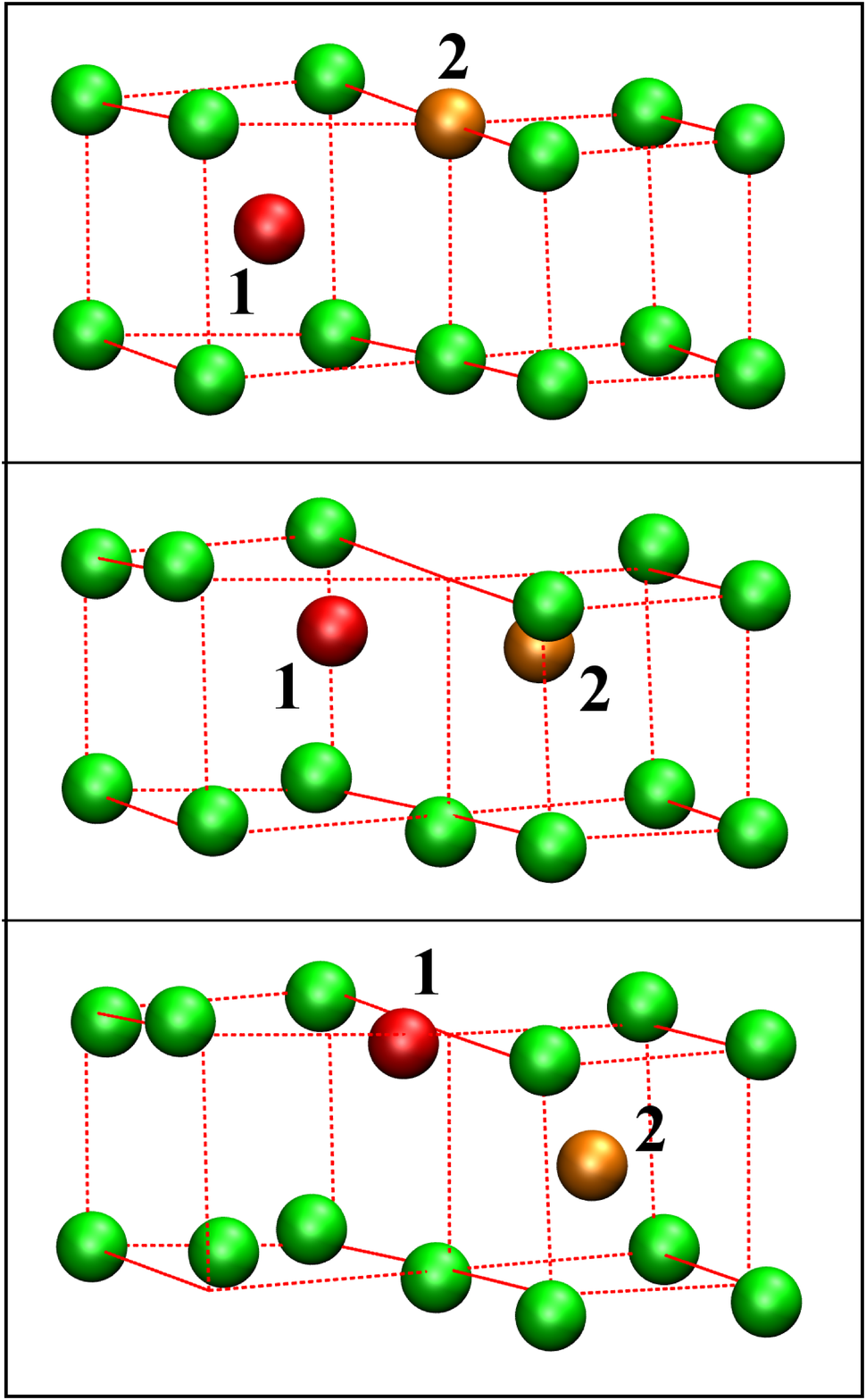}
        \caption{Diffusion exchange mechanism for Li interstitial. In the picture the initial, transition and final states are shown.}
        \label{fig:lidiff}
\end{figure*}
\clearpage

\section*{Debye-Waller factor}
We calculated the Debye-Waller factor for each species as a function of the temperature from harmonic phonons.
The mean square displacement as a function of the temperature in harmonic approximation is given by
\begin{equation}
\langle u_{\alpha}^2 \rangle = \frac{1}{N_{\alpha}} \sum_{m,i} \frac{\hbar}{\omega_m} %
                      \frac{|\mathbf{e}(m,i)|^2}{M_{\alpha}} %
                      \left[ n_B\left(\frac{\hbar \omega_m}{k_BT}\right)+\frac{1}{2}\right]
\end{equation}
were $\alpha$ runs over the atomic species, $M_{\alpha}$ is the mass of $\alpha\textrm{th}$ species,
$i$ runs over $N_{\alpha}$ atoms of species $\alpha$, while $\omega_m$ and $\mathbf{e}(m,i)$ are the 
frequency and the eigenvector of the $m$th harmonic phonon.
\begin{figure*}[!h]
        \includegraphics[width=0.55\textwidth]{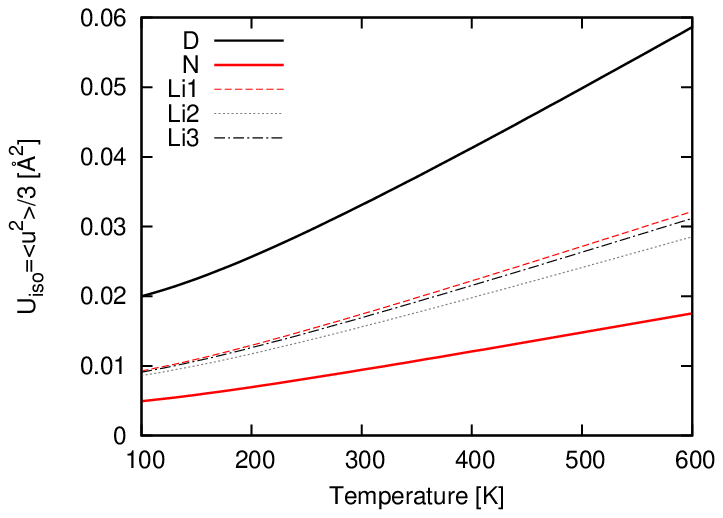}
        \caption{Debye-Waller factor, $U_{iso}=1/3 \langle u_{\alpha}^2 \rangle $, 
	for D, N, Li1, Li2, Li3 as a function of the temperature.}
        \label{fig:s1}
\end{figure*}
\newpage
\section*{Li$_2$NH X-ray diffraction pattern}
\begin{figure*}[!h]
\label{fig:xray}
\includegraphics[width=0.65\textwidth]{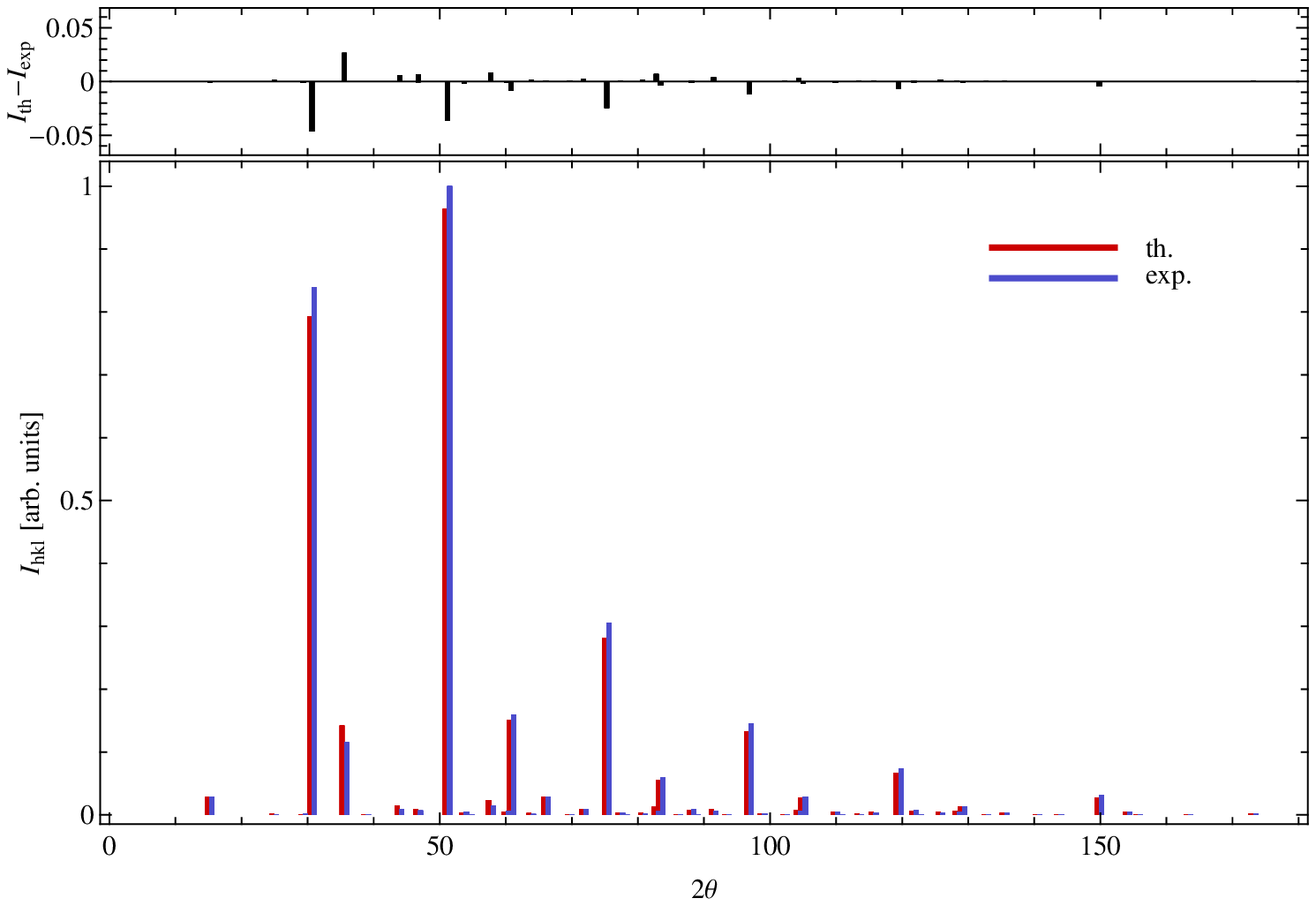}
\caption{Bragg reflections intensities computed the average positions and occupations reported
in Table~1, compared with those obtained using experimental values. No instrument-related corrections have been applied
and the same Debye-Waller factor ($U_{iso}=0.05$~\AA$^2$) has been used for all atomic species.}

\end{figure*}

We show in Fig.~\ref{fig:xray} a comparison of the calculated x-ray diffraction pattern
using both experimental~\cite{balogh} and our theoretical positions. 

\section*{Configurational entropy of tetrahedral clusters}
To compute the change in computational entropy due to the breaking of tetrahedra we considered
that the number of tetrahedra $N_T$ is related to the number of Li4 and Li2 sites by 
$N_{Li4}=6N_T$ and $N_{Li2}=3N_T$.
The change in free energy due to the breaking of a fraction $x$ of tetrahedra is given by
\begin{equation}
\frac{F}{N_T} = x \Delta \textrm{E}_0 - \frac{k_B T}{N_T} \left[ ln \binom{ 3N_T }{N_T(1-x)} - ln \binom{ 3N_T }{N_T} \right] %
- \frac{k_B T}{N_T} \, ln \binom{ N_T(2+4x) }{3x N_T} \nonumber
\end{equation}
which yields an equilibrium concentration of 
\begin{equation}
x \sim \sqrt[3]{4}/3\,exp[-\Delta \textrm{E}_0/3 k_B T] . \nonumber
\end{equation}


\begin{thebibliography}{10}%
\makeatletter
\providecommand \@ifxundefined [1]{%
 \ifx #1\undefined \expandafter \@firstoftwo
 \else \expandafter \@secondoftwo
\fi
}%
\providecommand \@ifnum [1]{%
 \ifnum #1\expandafter \@firstoftwo
 \else \expandafter \@secondoftwo
\fi
}%
\providecommand \enquote [1]{``#1''}%
\providecommand \bibnamefont  [1]{#1}%
\providecommand \bibfnamefont [1]{#1}%
\providecommand \citenamefont [1]{#1}%
\providecommand\href[0]{\@sanitize\@href}%
\providecommand\@href[1]{\endgroup\@@startlink{#1}\endgroup\@@href}%
\providecommand\@@href[1]{#1\@@endlink}%
\providecommand \@sanitize [0]{\begingroup\catcode`\&12\catcode`\#12\relax}%
\@ifxundefined \pdfoutput {\@firstoftwo}{%
 \@ifnum{\z@=\pdfoutput}{\@firstoftwo}{\@secondoftwo}%
}{%
 \providecommand\@@startlink[1]{\leavevmode}%
 \providecommand\@@endlink[0]{}%
}{%
 \providecommand\@@startlink[1]{%
  \leavevmode
  \pdfstartlink
   attr{/Border[0 0 1 ]/H/I/C[0 1 1]}%
   user{/Subtype/Link/A<</Type/Action/S/URI/URI(#1)>>}%
  \relax
 }%
 \providecommand\@@endlink[0]{\pdfendlink}%
}%
\providecommand \url  [0]{\begingroup\@sanitize \@url }%
\providecommand \@url [1]{\endgroup\@href {#1}{\urlprefix}}%
\providecommand \urlprefix [0]{URL }%
\providecommand \Eprint[0]{\href }%
\@ifxundefined \urlstyle {%
  \providecommand \doi [1]{doi:\discretionary{}{}{}#1}%
}{%
  \providecommand \doi [0]{doi:\discretionary{}{}{}\begingroup
  \urlstyle{rm}\Url }%
}%
\providecommand \doibase [0]{http://dx.doi.org/}%
\providecommand \Doi[1]{\href{\doibase#1}}%
\providecommand \bibAnnote [3]{%
  \BibitemShut{#1}%
  \begin{quotation}\noindent
    \textsc{Key:}\ #2\\\textsc{Annotation:}\ #3%
  \end{quotation}%
}%
\providecommand \bibAnnoteFile [2]{%
  \IfFileExists{#2}{\bibAnnote {#1} {#2} {\input{#2}}}{}%
}%
\providecommand \typeout [0]{\immediate \write \m@ne }%
\providecommand \selectlanguage [0]{\@gobble}%
\providecommand \bibinfo [0]{\@secondoftwo}%
\providecommand \bibfield [0]{\@secondoftwo}%
\providecommand \translation [1]{[#1]}%
\providecommand \BibitemOpen[0]{}%
\providecommand \bibitemStop [0]{}%
\providecommand \bibitemNoStop [0]{.\EOS\space}%
\providecommand \EOS [0]{\spacefactor3000\relax}%
\providecommand \BibitemShut [1]{\csname bibitem#1\endcsname}%
%</preamble>
\bibitem{ping1}%
  \BibitemOpen
  \bibfield{author}{%
  \bibinfo {author} {\bibnamefont{{P. Chen, Z. Xiong, J. Luo, J. Lin and K. L.
  Tan}}},\ }%
  \bibfield{journal}{%
  \bibinfo {journal} {{Nature}}\ }%
  \textbf{\bibinfo {volume} {{420}}},\ \bibinfo {pages} {302} (\bibinfo {year}
  {2002})%
  \bibAnnoteFile{NoStop}{ping1}%
\bibitem{ping2}%
  \BibitemOpen
  \bibfield{author}{%
  \bibinfo {author} {\bibnamefont{{P. Chen, Z. Xiong, J. Luo, J. Lin, and K. L.
  Tan}}},\ }%
  \bibfield{journal}{%
  \bibinfo {journal} {{J. Phys. Chem. B}}\ }%
  \textbf{\bibinfo {volume} {{107}}},\ \bibinfo {pages} {10967} (\bibinfo
  {year} {2003})%
  \bibAnnoteFile{NoStop}{ping2}%
\bibitem{gregory}%
  \BibitemOpen
  \bibfield{author}{%
  \bibinfo {author} {\bibnamefont{{D. H. Gregory}}},\ }%
  \bibfield{journal}{%
  \bibinfo {journal} {{J. Mat. Chem.}}\ }%
  \textbf{\bibinfo {volume} {{18}}},\ \bibinfo {pages} {1221} (\bibinfo {year}
  {2008})%
  \bibAnnoteFile{NoStop}{gregory}%
\bibitem{mtoday}%
  \BibitemOpen
  \bibfield{author}{%
  \bibinfo {author} {\bibnamefont{{P. Chen and M. Zhu}}},\ }%
  \bibfield{journal}{%
  \bibinfo {journal} {{Materials Today}}\ }%
  \textbf{\bibinfo {volume} {{11}}},\ \bibinfo {pages} {36} (\bibinfo {year}
  {2008})%
  \bibAnnoteFile{NoStop}{mtoday}%
\bibitem{orimo}%
  \BibitemOpen
  \bibfield{author}{%
  \bibinfo {author} {\bibnamefont{{S. Orimo, Y. Nakamori, J. F. Elise, A.
  Z\"uttel, and C. M. Jensen}}},\ }%
  \bibfield{journal}{%
  \bibinfo {journal} {{Chem. Rev.}}\ }%
  \textbf{\bibinfo {volume} {{107}}},\ \bibinfo {pages} {4111} (\bibinfo {year}
  {2007})%
  \bibAnnoteFile{NoStop}{orimo}%
\bibitem{yukawa}%
  \BibitemOpen
  \bibfield{author}{%
  \bibinfo {author} {\bibfnamefont{H.}~\bibnamefont{Yukawa}}\ and\ \bibinfo
  {author} {\bibnamefont{{\em et al.}}},\ }%
  \bibfield{journal}{%
  \bibinfo {journal} {J. Alloys Compd.}\ }%
  \textbf{\bibinfo {volume} {242}},\ \bibinfo {pages} {446} (\bibinfo {year}
  {2007})%
  \bibAnnoteFile{NoStop}{yukawa}%
\bibitem{wood-marz09prl}%
  \BibitemOpen
  \bibfield{author}{%
  \bibinfo {author} {\bibfnamefont{B.~C.}\ \bibnamefont{Wood}}\ and\ \bibinfo
  {author} {\bibfnamefont{N.}~\bibnamefont{Marzari}},\ }%
  \bibfield{journal}{%
  \bibinfo {journal} {Phys. Rev. Lett.}\ }%
  \textbf{\bibinfo {volume} {103}},\ \bibinfo {pages} {185901} (\bibinfo {year}
  {2009})%
  \bibAnnoteFile{NoStop}{wood-marz09prl}%
\bibitem{imideOD1}%
  \BibitemOpen
  \bibfield{author}{%
  \bibinfo {author} {\bibnamefont{{R. A. Forman}}},\ }%
  \bibfield{journal}{%
  \bibinfo {journal} {{J. Chem. Phys.}}\ }%
  \textbf{\bibinfo {volume} {{55}}},\ \bibinfo {pages} {1987} (\bibinfo {year}
  {1971})%
  \bibAnnoteFile{NoStop}{imideOD1}%
\bibitem{imideOD2}%
  \BibitemOpen
  \bibfield{author}{%
  \bibinfo {author} {\bibnamefont{{P.J. Haigh, R. A. Forman and R. C.
  Frish}}},\ }%
  \bibfield{journal}{%
  \bibinfo {journal} {{J. Chem. Phys.}}\ }%
  \textbf{\bibinfo {volume} {{45}}},\ \bibinfo {pages} {812} (\bibinfo {year}
  {1966})%
  \bibAnnoteFile{NoStop}{imideOD2}%
\bibitem{balogh}%
  \BibitemOpen
  \bibfield{author}{%
  \bibinfo {author} {\bibnamefont{{M. P. Balogh, C. Y. Jones, J. F. Herbst, L.
  G. Hector Jr and M. Kundrat}}},\ }%
  \bibfield{journal}{%
  \bibinfo {journal} {{J. Alloys and Comp.}}\ }%
  \textbf{\bibinfo {volume} {{420}}},\ \bibinfo {pages} {326} (\bibinfo {year}
  {2006})%
  \bibAnnoteFile{NoStop}{balogh}%
\bibitem{rijss}%
  \BibitemOpen
  \bibfield{author}{%
  \bibinfo {author} {\bibfnamefont{J.}~\bibnamefont{Rijssenbeek}}, \bibinfo
  {author} {\bibfnamefont{Y.}~\bibnamefont{Gao}}, \bibinfo {author}
  {\bibfnamefont{J.}~\bibnamefont{Hanson}}, \bibinfo {author}
  {\bibfnamefont{Q.}~\bibnamefont{Huang}}, \bibinfo {author}
  {\bibfnamefont{C.}~\bibnamefont{Jones}},\ and\ \bibinfo {author}
  {\bibfnamefont{B.}~\bibnamefont{Toby}},\ }%
  \bibfield{journal}{%
  \bibinfo {journal} {Journal of Alloys and Compounds}\ }%
  \textbf{\bibinfo {volume} {454}},\ \bibinfo {pages} {233} (\bibinfo {year}
  {2008})%
  \bibAnnoteFile{NoStop}{rijss}%
\bibitem{herbst}%
  \BibitemOpen
  \bibfield{author}{%
  \bibinfo {author} {\bibnamefont{{J. F. Herbst and L. G. Hector Jr}}},\ }%
  \bibfield{journal}{%
  \bibinfo {journal} {{Phys. Rev. B}}\ }%
  \textbf{\bibinfo {volume} {{72}}},\ \bibinfo {pages} {125120} (\bibinfo
  {year} {2005})%
  \bibAnnoteFile{NoStop}{herbst}%
\bibitem{hector}%
  \BibitemOpen
  \bibfield{author}{%
  \bibinfo {author} {\bibnamefont{{L. G. Hector Jr and J. F. Herbst}}},\ }%
  \bibfield{journal}{%
  \bibinfo {journal} {{J. Phys. Cond. Matt.}}\ }%
  \textbf{\bibinfo {volume} {{20}}},\ \bibinfo {pages} {064229} (\bibinfo
  {year} {2008})%
  \bibAnnoteFile{NoStop}{hector}%
\bibitem{wolverton}%
  \BibitemOpen
  \bibfield{author}{%
  \bibinfo {author} {\bibnamefont{{B. Magyari-K\"ope, V. Ozolins, and C.
  Wolverton}}},\ }%
  \bibfield{journal}{%
  \bibinfo {journal} {{Phys. Rev. B}}\ }%
  \textbf{\bibinfo {volume} {{73}}},\ \bibinfo {pages} {20101(R)} (\bibinfo
  {year} {2006})%
  \bibAnnoteFile{NoStop}{wolverton}%
\bibitem{ceder}%
  \BibitemOpen
  \bibfield{author}{%
  \bibinfo {author} {\bibnamefont{{T. Mueller and G. Ceder}}},\ }%
  \bibfield{journal}{%
  \bibinfo {journal} {{Phys. Rev. B}}\ }%
  \textbf{\bibinfo {volume} {{74}}},\ \bibinfo {pages} {134104} (\bibinfo
  {year} {2006})%
  \bibAnnoteFile{NoStop}{ceder}%
\bibitem{mice+2010-jpcc}%
  \BibitemOpen
  \bibfield{author}{%
  \bibinfo {author} {\bibfnamefont{G.}~\bibnamefont{Miceli}}, \bibinfo {author}
  {\bibfnamefont{C.}~\bibnamefont{Cucinotta}}, \bibinfo {author}
  {\bibfnamefont{M.}~\bibnamefont{Bernasconi}},\ and\ \bibinfo {author}
  {\bibfnamefont{M.}~\bibnamefont{Parrinello}},\ }%
  \bibfield{journal}{%
  \bibinfo {journal} {J. Phys. Chem. C},\ \bibinfo {pages} {302}}%
   (\bibinfo {year} {2010})%
  \bibAnnoteFile{NoStop}{mice+2010-jpcc}%
\bibitem{sebastiani}%
  \BibitemOpen
  \bibfield{author}{%
  \bibinfo {author} {\bibfnamefont{G.~A.}\ \bibnamefont{Ludueña}}, \bibinfo
  {author} {\bibfnamefont{M.}~\bibnamefont{Wegner}}, \bibinfo {author}
  {\bibfnamefont{L.}~\bibnamefont{Bjalie}},\ and\ \bibinfo {author}
  {\bibfnamefont{D.}~\bibnamefont{Sebastiani}},\ }%
  \bibfield{journal}{%
  \bibinfo {journal} {Chem. Phys. Chem.}\ }%
  \textbf{\bibinfo {volume} {11}},\ \bibinfo {pages} {2353} (\bibinfo {year}
  {2010})%
  \bibAnnoteFile{NoStop}{sebastiani}%
\bibitem{PBE}%
  \BibitemOpen
  \bibfield{author}{%
  \bibinfo {author} {\bibnamefont{{J. P. Perdew, K. Burke, and M.
  Ernzerhof}}},\ }%
  \bibfield{journal}{%
  \bibinfo {journal} {{Phys. Rev. Lett.}}\ }%
  \textbf{\bibinfo {volume} {{77}}},\ \bibinfo {pages} {3865} (\bibinfo {year}
  {1996})%
  \bibAnnoteFile{NoStop}{PBE}%
\bibitem{CPMD}%
  \BibitemOpen
  \enquote{\bibinfo {title} {{CPMD}},}\ \bibinfo {note}
  {{http://www.cpmd.org/}}%
  \bibAnnoteFile{NoStop}{CPMD}%
\bibitem{vanderbilt}%
  \BibitemOpen
  \bibfield{author}{%
  \bibinfo {author} {\bibnamefont{{D. Vanderbilt}}},\ }%
  \bibfield{journal}{%
  \bibinfo {journal} {{Phys. Rev. B}}\ }%
  \textbf{\bibinfo {volume} {{41}}},\ \bibinfo {pages} {7892} (\bibinfo {year}
  {1990})%
  \bibAnnoteFile{NoStop}{vanderbilt}%
\bibitem{GTH1}%
  \BibitemOpen
  \bibfield{author}{%
  \bibinfo {author} {\bibnamefont{{S. Goedecker, M. Teter, and J. Hutter}}},\
  }%
  \bibfield{journal}{%
  \bibinfo {journal} {{Phys. Rev. B}}\ }%
  \textbf{\bibinfo {volume} {{54}}},\ \bibinfo {pages} {1703} (\bibinfo {year}
  {1996})%
  \bibAnnoteFile{NoStop}{GTH1}%
\bibitem{colored-jctc}%
  \BibitemOpen
  \bibfield{author}{%
  \bibinfo {author} {\bibfnamefont{M.}~\bibnamefont{Ceriotti}}, \bibinfo
  {author} {\bibfnamefont{G.}~\bibnamefont{Bussi}},\ and\ \bibinfo {author}
  {\bibfnamefont{M.}~\bibnamefont{Parrinello}},\ }%
  \bibfield{journal}{%
  \bibinfo {journal} {J. of Chem. Theory and Comput.}\ }%
  \textbf{\bibinfo {volume} {6}},\ \bibinfo {pages} {1170} (\bibinfo {year}
  {2010})%
  \bibAnnoteFile{NoStop}{colored-jctc}%
\bibitem{Quantum-espresso}%
  \BibitemOpen
  \bibfield{author}{%
  \bibinfo {author} {\bibfnamefont{P.}~\bibnamefont{{Giannozzi {\em et
  al.}}}},\ }%
  \enquote{\bibinfo {title} {Quantum-espresso},}\ \bibinfo {note}
  {{http://www.quantum-expresso.org, http://www.pwscf.org}}%
  \bibAnnoteFile{NoStop}{Quantum-espresso}%
\bibitem{NEB2}%
  \BibitemOpen
  \bibfield{author}{%
  \bibinfo {author} {\bibnamefont{{G. Henkelman, B. P. Uberuaga, and H.
  J\'onsson}}},\ }%
  \bibfield{journal}{%
  \bibinfo {journal} {{J. Chem. Phys.}}\ }%
  \textbf{\bibinfo {volume} {{113}}},\ \bibinfo {pages} {9901} (\bibinfo {year}
  {2000})%
  \bibAnnoteFile{NoStop}{NEB2}%
\bibitem{sosso}%
  \BibitemOpen
  \bibfield{author}{%
  \bibinfo {author} {\bibfnamefont{G.}~\bibnamefont{Sosso}}, \bibinfo {author}
  {\bibfnamefont{S.}~\bibnamefont{Caravati}}, \bibinfo {author}
  {\bibfnamefont{C.}~\bibnamefont{Gatti}}, \bibinfo {author}
  {\bibfnamefont{S.}~\bibnamefont{Assoni}},\ and\ \bibinfo {author}
  {\bibfnamefont{M.}~\bibnamefont{Bernasconi}},\ }%
  \bibfield{journal}{%
  \bibinfo {journal} {J. Phys. Cond. Matt.}\ }%
  \textbf{\bibinfo {volume} {21}},\ \bibinfo {pages} {245401} (\bibinfo {year}
  {2009})%
  \bibAnnoteFile{NoStop}{sosso}%
\end{thebibliography}
\end{document}